\documentstyle[aps,twocolumn]{revtex}

\input psfig.sty

\begin{document}
\draft
\title{Complete Characterization of a Quantum Process: \\
the Two--Bit Quantum Gate}
\author{J.~ F.~ Poyatos, J.~ I.~ Cirac}
\address{Departamento de Fisica Aplicada, Universidad de
Castilla--La Mancha, 13071 Ciudad Real, Spain}
\author{P.~ Zoller}
\address{Institut f{\"u}r Theoretische Physik,
Universit{\"a}t Innsbruck, A--6020
Innsbruck, Austria}
\date{\today}
\maketitle

\begin{abstract}
We show how to fully characterize a quantum process in an open quantum
system. We particularize the procedure to the case of a universal two--qubit
gate in a quantum computer. We illustrate the method with a numerical
simulation of a quantum gate in the ion trap quantum computer.
\end{abstract}

\pacs{PACS Nos. 03.65.Bz, 42.50.Lc, 42.50.Wm}

\narrowtext

Recently there has been a growing interest in ``quantum tomography'',
i.e.~in the complete characterization of the {\it state of a quantum system }
represented by a density operator $\hat{\rho}$. Quantum tomography of an
unknown quantum state (that can be repeatedly prepared) \cite{Le95} consists
of finding an appropriate sequence of measurements which allows one to
determine the complete density operator $\hat{\rho}$ (for experimental
implementations and theoretical schemes in quantum optics see \cite{Le95}).
In this letter we will show how to completely characterize a {\it physical
process in an open quantum system}. More specifically, suppose that a given
quantum dynamics ${\cal E}$ transforms input states $\rho _{{\rm in}}$ into
output states $\rho _{{\rm out}}$, i.e. 
\begin{equation}
\hat{\rho}_{{\rm in}}\stackrel{\cal E}{\longrightarrow }\hat{\rho}_{{\rm out}
}={\cal E}[\hat{\rho}_{{\rm in}}]  \label{trafo}
\end{equation}
with ${\cal E}$ a linear mapping. Our aim is to characterize the process 
${\cal E}$, given as a ``black box'', by a sequence of measurements in such a
way that it is possible to predict what the output state will be for any
input state.

The particular problem that we will analyze after developing a general formalism
is the characterization of the two-bit universal quantum gate 
for quantum computing \cite{Di95}. A quantum
computer consists of $n$ two--level atoms with atomic states  $|0\rangle
_i,|1\rangle _i$ ($i=1,\ldots ,n$) representing the quantum bits (qubits).
States of the quantum computer are $n$-atom entangled states in the product
Hilbert space $|\psi \rangle \in {\cal H}=\prod_i\otimes {\cal H}_2(i)$ with 
${\cal H}_2(i)=\left\{ |0\rangle _i,|1\rangle _i\right\} $. Quantum
computations correspond to physical processes $|\psi _{{\rm out}}\rangle =
\hat{U}|\psi _{{\rm in}}\rangle $ where a given input state is mapped to an
output state by a unitary transformation $\hat{U}$. This can be carried out as a
sequence of elementary steps (quantum gates) involving operations on a few
qubits. It has been shown that any computation can be decomposed into
single-bit gates, and a universal two-bit gate which involves an
entanglement operation on two qubits\cite{Di95}. In reality, due to the
presence of decoherence and experimental imperfections, these gates (and
therefore any computation) will not be ideal. In present experiments related to
quantum computing based on both laser cooled trapped ions \cite{Mo95} and
atoms in cavities\cite{Tu95}, the difficult part is the two--qubit
gate, since it requires an interaction between the two two--level systems
via an auxiliary system (phonons or photons) which leads to decoherence. In
view of this fact, we wish to develop a procedure for characterizing a
two--qubit gate, i.e. characterize a physical process ${\cal E}$ involving
entanglement of two qubits 1 and 2 in the state space ${\cal H}_2(1)\otimes 
{\cal H}_2(2)$. Below we will show how to implement this using only 
{\it product states as inputs}, and {\it single qubit measurements on the
outputs }(assuming that single bit preparations and operations can be
performed reliably ). We avoid utilizing any interaction (entanglement)
between the qubits which would be required to prepare Bell state inputs and
perform Bell measurements since otherwise the decoherence and errors induced
by the measurement itself would distort the characterization of ${\cal E}$.
Furthermore, we will introduce four global parameters to characterize the
action of the quantum gate: the ``Gate Fidelity'' (${\cal F}$), the
``Gate Purity'' (${\cal P}$), the ``Quantum Degree'' 
(${\cal Q}$), and the ``Entanglement Capability'' (${\cal E}$).
These four parameters can be
calculated once the physical process is completely determined. To illustrate
the procedure of measuring ${\cal E}$, we will analyze below simulation data
from a model of a two-bit quantum gate in an ion trap quantum computer\cite
{Ci95}.

To develop the general formalism, let us consider an experiment in which
a quantum system undergoes a physical process ${\cal E}$. We assume that the
system is initially prepared in the pure state 
\begin{equation}
|\Psi _{{\rm in}}\rangle =\sum_{i=0}^Nc_i|i\rangle \in {\cal H}_{{\rm in}},
\label{init}
\end{equation}
where $|0\rangle ,|1\rangle ,\ldots ,|N\rangle $ are orthogonal states
spanning the Hilbert space of (allowed) input states ${\cal H}_{{\rm in}}$
with dimension $N+1$, a subspace of the system Hilbert space ${\cal H}_{{\rm 
S}}$. We will denote by $\rho_E=\sum_\alpha \omega_\alpha |E^\alpha)(E^\alpha|$
the initial state of the
environment, i.e., of the other degrees of freedom that will be coupled to
our system. In its
most general form, the physical process will perform a transformation
defined by 
\begin{equation} \label{two}
|i\rangle |E^\alpha)\stackrel{\cal E}{\longrightarrow }\sum_{j=0}^M|j\rangle
|E^\alpha_{ij}),\quad \text{(}i=0,\ldots ,N\text{)} 
\end{equation}
where in the sum we have taken into account the possible presence of other
system states that might be populated during the interaction (i.e., $M\ge N$
in general; see also below for the case of a quantum gate). The states 
$|E^\alpha_{ij})$ are unnormalized states of the environment. 
Combining (\ref{init})
and (\ref{two}), and tracing over the environment degrees of freedom we get
the following reduced system density operator, 
\begin{equation}
\hat{\rho}_{{\rm out}}=\sum_{i,i^{\prime }=0}^Nc_i\left[ c_{i^{\prime
}}\right] ^{*}\hat{R}_{i^{\prime }i}  \label{evol}
\end{equation}
in the space of output states, ${\cal H}_{{\rm out}}=\{|0\rangle ,\ldots
,|M\rangle \}$, and where 
\begin{equation}
\hat{R}_{i^{\prime }i}\equiv \sum_{j,j^{\prime }=0}^M
|j\rangle \langle j^{\prime }|
\sum_\alpha \omega_\alpha 
(E^\alpha_{i^{\prime }j^{\prime }}|E^\alpha_{ij})
\quad \text{(}i,i
\acute{}=0,\ldots ,N\text{)}
\end{equation}
are system operators that do not depend on the initial state. From the knowledge
of these operators one can predict the final density operator $\hat{\rho}_{
{\rm out}}$ for any input state $\rho _{{\rm in}}$. For a mixed initial
state, we diagonalize $\hat{\rho}_{{\rm in}}=\sum_{n}p_n|\Psi_{\rm
in}^{(n)}\rangle \langle \Psi_{\rm in}^{(n)}|$ and use the fact that the 
evolution ${\cal E}$ is linear, which leads to 
$\hat{\rho}_{{\rm out}}=
\sum_{n}p_n{\cal E}(|\Psi_{\rm in}^{(n)}\rangle \langle \Psi_{\rm in}^{(n)}|)$. 
Thus, the problem of fully
characterizing the physical process ${\cal E}$ on the system is reduced to
finding the $(N+1)^2$ ``transfer operators'' $\hat{R}_{i^{\prime }i}$ in $
{\cal H}_{{\rm out}}$. They fulfill ${\rm Tr}\{\hat{R}_{i^{\prime
}i}\}=\delta _{i^{\prime }i}$ and $(\hat{R}_{i^{\prime }i})^{\dagger }=\hat{R
}_{ii^{\prime }}$.

To develop a procedure to measure these operators we define two vectorial
operators 
\begin{mathletters}
\begin{eqnarray}
\vec{\rho}_{{\rm out}} &\equiv &\left\{ \hat{\rho}_{{\rm out}}^{(1)},\hat{
\rho}_{{\rm out}}^{(2)},\ldots ,\hat{\rho}_{{\rm out}}^{(N+1)^2}\right\} , \\
\vec{R} &\equiv &\left\{ \hat{R}^1,\hat{R}^2,\ldots ,\hat{R}
^{(N+1)^2}\right\} 
\end{eqnarray}
Here, the components of $\vec{\rho}_{out}$ are a set of output operators
corresponding to $(N+1)^2$ different initial inputs of the form (\ref{init})
with coefficients $c_i^{(k)}$ [$k=1,\ldots ,(N+1)^2$]. All these density
operators can be fully characterized using standard quantum tomography
methods \cite{Le95}. On the other hand, the components of the vector $\vec{R}
$, are defined according to $\hat{R}_q=\hat{R}_{i^{\prime }i}$, with 
$q=(N+1)i^{\prime }+i+1$. In view of (\ref{evol}), these two vectorial
operators are related by a set of linear equations, written in a matrix form
as $\vec{\rho}_{{\rm out}}={\cal M}\vec{R}$, where ${\cal M}$ is a (c--number) matrix
whose elements are defined as ${\cal M}_{kq}=c_i^{(k)}[c_{i^{\prime
}}^{(k)}]^{*}$. The problem of obtaining the transfer operators 
$\vec{R}$ thus reduces to finding a set of initial states of the system, such
that the matrix ${\cal M}$ is not singular. 
In that case we will have $\vec{R}={\cal M}^{-1}%
\vec{\rho}_{{\rm out}}$, which solves the problem. Next, we prove that such
an invertible matrix ${\cal M}$ exists by explicit construction. One can simply
choose the initial states as follows: 
\end{mathletters}
\begin{equation}
c_i^{(k)}=\left\{ 
\begin{array}{ll}
\frac 1{\sqrt{2}}(\delta _{ik_1}+\delta _{ik_2}) & {\rm if}\;k_1>k_2, \\ 
\delta _{ik_1} & {\rm if}\;k_1=k_2, \\ 
\frac 1{\sqrt{2}}(\delta _{ik_1}+i\delta _{ik_2}) & {\rm if}\;k_1<k_2,
\end{array}
\right.   \label{inst}
\end{equation}
with $k=(N+1)k_1+k_2$.

Let us now apply the above procedure to a universal two--qubit gate. Now,
the system is composed of two two--level subsystems $1$ and $2$ of levels 
$|0\rangle _{1,2}$ and $|1\rangle _{1,2}$ each. We can define a set of
orthogonal states $|i\rangle =|i_1\rangle _1|i_2\rangle _1$ (with $i=2i_1+i_2
$, and $i_1,i_2=0,1$) and write the initial states as in (\ref{init}) (with 
dimension $N+1=4$).
The quantum tomography of the output states can be carried out
following the lines proposed by Wootters \cite{Wo87}: one writes the output
density operator as 
\begin{equation}
\label{Woot}
\hat{\rho}_{{\rm out}}=\sum_{q=0}^{15}\lambda _q\hat{A}_q,
\end{equation}
where $\hat{A}_q=\hat{\sigma} _{q_1}^1
\otimes \hat{\sigma}_{q_2}^2$ ($q=4q_1+q_2$),with 
$\hat{\sigma}_{q_i}^a=\{\hat{1}^a,\hat{\sigma}_x^a,\hat{\sigma}_y^a,
\hat{\sigma}_z^a\}$ and $a=1,2$ refers
to the first and second qubit, respectively. By measuring the observables 
$\hat{A}_q$, one can determine the coefficients $\lambda _q$, given that 
$\lambda _q={\rm Tr}[\hat{\rho}_{{\rm out}}\hat{A}_q]/4$ \cite{note1}. Note that all
these measurements do not require any interaction between the qubits (Bell
state measurements), that is, for all these measurements the two qubits can
be measured independently, without the application of another two--qubit
gate. This is needed since otherwise the measurement procedure would involve
errors that could not be separated from the gate itself. However, some of the
16 initial states given above in order to make up the matrix ${\cal M}$ [c.f. 
Eq.~(\ref{inst})] are entangled states. Their preparation would involve the
application of a two--qubit gate which would also lead to uncontrollable
errors. Fortunately, there are other sets of initial states that are
unentangled for which the matrix ${\cal M}$ is invertible. An example are the 16
product states $|\psi _a\rangle _1|\psi _b\rangle _2$ ($a,b=1,\ldots ,4$),
where
\begin{equation}
\begin{array}{ll}
|\psi _1\rangle =|0\rangle , & |\psi _3\rangle =\frac 1{\sqrt{2}}(|0\rangle
+|1\rangle ) \\ 
\,|\psi _2\rangle =|1\rangle ,\, & |\psi _4\rangle =\frac 1{\sqrt{2}}
(|0\rangle +i|1\rangle ).
\end{array}
\end{equation}

In order to illustrate this procedure, we have studied a two--qubit gate in
the ion trap quantum computer model \cite{Ci95}. We have considered two ions
in a linear ion trap interacting with two lasers. Let us denote by 
$|g\rangle _n\equiv |0\rangle _n$ and $|e\rangle _n\equiv |1\rangle _n$ two
internal states of the n--th ion, and by $|e^{\prime }\rangle _n$ an
auxiliary internal state. As we have shown in Ref.~\cite{Ci95}, the universal
two--qubit gate defined by 
\begin{equation}
|\epsilon _1\rangle _1|\epsilon _2\rangle _2\rightarrow (-1)^{\epsilon
_1\epsilon _2}|\epsilon _1\rangle _1|\epsilon _2\rangle _2,\quad \quad
(\epsilon _{1,2}=0,1)  \label{gate}
\end{equation}
can be implemented in three steps: (i) Apply a $\pi $ laser pulse to the
lower motional sideband corresponding to the transition $|g\rangle
_1\rightarrow |e\rangle _1$ of the first ion; (ii) Apply a $2\pi $ laser
pulse to the lower motional sideband of the transition $|g\rangle
_2\rightarrow |e^{\prime }\rangle _2$ of the second ion; (iii) as (i). By
lower motional sideband we mean that the laser frequency has to be equal to
the corresponding internal transition frequency minus the trap frequency, in
order to excite a center of mass phonon only. The interaction of the two ions and
the laser is given by the following Hamiltonian \cite{note2} 
\begin{eqnarray*}
H &=&-\Delta _1|e\rangle _1{}_1\!\langle e|-\Delta _2|e^{\prime }\rangle
_2{}_2\!\langle e^{\prime }|+\nu a_{{\rm cm}}^{\dagger }a_{{\rm cm}}+\sqrt{3}
\nu a_{{\rm r}}^{\dagger }a_{{\rm r}} \\
&&+\frac{\Omega _1(t)}2[|e\rangle _1{}_1\!\langle g|e^{-i\eta _{{\rm cm}}
(a_{{\rm cm}}+a_{{\rm cm}}^{\dagger })}e^{-i\eta _{{\rm r}}(a_{{\rm r}}+
a_{{\rm r}}^{\dagger })}+{\rm H.c.}] \\
&&+\frac{\Omega _2(t)}2[|e^{\prime }\rangle _2{}_2\!\langle g|e^{-i
\eta_{{\rm cm}}(a_{{\rm cm}}+a_{{\rm cm}}^{\dagger })}e^{i\eta _{{\rm r}}
(a_{{\rm r}}+a_{{\rm r}}^{\dagger })}+{\rm H.c.}].
\end{eqnarray*}
Here, $\Delta _{1,2}$ and $\Omega _{1,2}$ are the laser detunings and Rabi
frequencies of the laser acting on each ion, respectively. The operators $a$
and $a^{\dagger }$ are annihilation and creation operators of the center of
mass (cm) and relative (r) motion mode, $\eta $ is the corresponding
Lamb--Dicke parameter, and $\nu $ is the trap frequency.

We have calculated numerically the evolution given by this Hamiltonian.
Following the steps mentioned above, and for Rabi frequencies much smaller
than the trap frequency this evolution corresponds to the gate (\ref{gate}) 
\cite{Ci95}. For finite Rabi frequencies, however, the result will not be
ideal. After the gate operation, there will remain some population in the
phonon modes, which would lead to decoherence. Moreover, there may remain
some population in the auxiliary state $|e^{\prime}\rangle_2$, and therefore
in this case we have $M=4>N=3$. With the numerical calculation we have
simulated the measurement of the operators $\hat{R}_{i^{\prime}i}$, that
fully characterize the evolution process. We emphasize that all the
information regarding this (non--ideal) gate is contained in these
operators. In Fig.~1 we have compared the ideal case of a perfect gate
[Fig.~1(a)] with the simulation results with realistic parameters
[Figs.~1(b--d)]. We have plotted the matrix elements of 16 operators $\hat{R}
_{i^{\prime}i}$ sorted according to ${\cal E}_{n,m}=\langle
j^{\prime}|\hat{R}_{i^{\prime}i}| j\rangle$, with $n=4i+j$ and 
$m=4i^{\prime}+j^{\prime}$. We have chosen a set of parameters close to
those planned in experiments: $\eta_{{\rm cm}}=\eta_{{\rm r}}=0.5$, 
$\Delta_{1}=\Delta_{2}=-\nu$, and Rabi frequencies 
$\Omega_1=\Omega_2=0.1,0.2\nu$
and $0.5$ [Figs.1(b,c,d)] \cite{note3}. As it is shown, for moderately small Rabi
frequencies the simulated results almost coincide with the ideal ones.

In order to elucidate to what extent a two--qubit gate implemented
experimentally in a particular model of quantum computer approaches to the
ideal one, we define a parameter, the ``Gate Fidelity'', as 
\[
{\cal F}=\overline{\langle \Psi _{{\rm in}}|\hat{U}^{\dagger }
\hat{\rho}_{{\rm out}}\hat{U}|\Psi _{{\rm in}}\rangle },
\]
where the overline indicates average over all possible input states 
$|\Psi _{{\rm in}}\rangle $, and $\hat{U}$ is the unitary operator corresponding to
the ideal gate. This parameter can be calculated once the full
characterization of the gate is performed using 
\[
{\cal F}=\frac 18\sum_{i=0}^3F_{ii}^{ii}+\frac 1{24}\sum_{i\neq
j}(F_{jj}^{ii}+F_{ij}^{ji}),
\]
where $F_{j^{\prime }j}^{i^{\prime }i}\equiv \langle j^{\prime }|
\hat{U}^{\dagger }\hat{R}_{i^{\prime }i}\hat{U}|j\rangle $. Obviously, a gate
fidelity close to one indicates that the gate was carried out almost ideally.

In a similar way we can define the ``Gate Purity'' ${\cal P}=\overline{{\rm Tr}
\{(\hat \rho _{{\rm out}})^2\}}$ that reflects the effects of decoherence on
the gate. ${\cal P}$ close to one indicates that the effects of decoherence
are negligible. It can be shown that 
\[
{\cal P}=\frac 18\sum_{i=0}^3{\rm Tr}\{(\hat{R}_{ii})^2\}+\frac 1{24}
\sum_{i\neq j}{\rm Tr}\{\hat{R}_{ii}\hat{R}_{i^{\prime }i^{\prime }}+\hat{R}
_{i^{\prime }i}\hat{R}_{ii^{\prime }}\}.
\]

In addition, the ``Quantum Degree of the Gate'' ${\cal Q}$ is defined as the
maximum value of the overlap between all possible output states that are
obtained starting from an unentangled state and all the maximally entangled
states, i.e. 
\[
{\cal Q}=\max_{\tilde{\rho}_{{\rm out}},|\Psi _{{\rm me}}\rangle }\langle
\Psi _{{\rm me}}|\tilde{\rho}_{{\rm out}}|\Psi _{{\rm me}}\rangle ,
\]
where $\tilde{\rho}_{{\rm out}}$ denote the output states corresponding to
unentangled input states $|\Psi _{{\rm in}}\rangle =|\psi _a\rangle _1|\psi
_b\rangle _2$, and $|\Psi _{{\rm me}}\rangle $ is a maximum entangled state
\cite{Pe96}. As it has been
shown, when the overlap between a density operator and a maximally entangled
state is larger than $(2+3\sqrt{2})/8\simeq 0.78$,
Clauser--Horne--Shimony--Holt (CHSH) inequalities are violated \cite{Be96}.
Finally, another useful parameter is the ``Entanglement Capability'' ${\cal E}$
\cite{Pe96}, given as the
smallest eigenvalue of the partial transposed density matrix $\hat\rho_{\rm out}$,
for unentangled inputs states. As it has been recently shown \cite{Pe96},
the negativity of this quantity is a necessary and sufficient condition
for non--separability of density operators of two spin--$1/2$ systems.
These quantities can be calculated numerically starting from the gate operators 
$\hat{R}_{i^{\prime }i}$ with a maximization/minimization procedure. 

In Fig.~2 we have plotted these four parameters as a function of the Rabi
frequency $\Omega\equiv\Omega_1=\Omega_2$ for three different Lamb--Dicke
parameters and the same parameters as in Fig.~1. As expected, the best
results are obtained for small Rabi frequencies. This is due to the fact
that transitions to undesired levels are suppressed. On the other
hand, a Lamb--Dicke parameter close to one also improves the results. This
is due to the fact that in contrast to Ref.~\cite{Ci95} we are considering
here traveling wave excitation, in which transitions that do not change the
phonon number can be excited by the laser. The ratio between the effective
Rabi frequencies of the lower sideband and these unwanted ones is
proportional to $\eta$ in the Lamb--Dicke limit. The figures also indicate
that the ${\cal F}$, ${\cal P}$, ${\cal Q}$, and ${\cal E}$ decay in a different
manner when the parameters deviate from the conditions of operation of the
ion trap quantum computer. 

Finally, we wish to mention another application of this procedure of
characterization of a quantum physical process. Let us consider a quantum
system coupled to a Markovian reservoir. In the case that the Hamiltonian is
time independent, one can describe the evolution of the reduced system
density operator $\hat{\rho}$ in terms of a master equation of the form 
$\dot{\hat{\rho}}={\cal L}\hat{\rho}$, where ${\cal L}$ is a Liouvillian
superoperator. The formal solution to this equation after a time $t$ is 
$\hat{\rho}_{{\rm out}}=e^{{\cal L}t}\hat{\rho}_{{\rm in}}$. The full
characterization of the physical process in this case would allow one to
``measure'' the Liouvillian ${\cal L}$, and therefore to determine the
master equation fulfilled by the system. Moreover, choosing different
interaction times one could check whether a given process is Markovian or
not.

In conclusion, we have shown how to perform the full characterization of a 
{\it physical process}. This requires the preparation of various initial
states, and the quantum tomography of the corresponding output states. We
have illustrated these results for the case of a quantum gate, and have
defined four parameters that give the quality of the gate. In addition, we
have presented numerical simulations for the ion trap quantum computer. The
complete characterization of the quantum gates is relevant both from an
experimental point of view to evaluate implementations of quantum gates in
the laboratory, as well as a theoretical tool to compare the expected
performance of specific quantum computer model systems.

We thank A. Barenco, J. Kimble, H. Mabuchi, T. Pellizzari, 
and R. Walser for discussions. 
This work was supported by the Austrian Science Foundation, ``Acciones 
Integradas'',
and the European TMR network
ERB4061PL95--1412. J. F. Poyatos acknowledges the ``J. C.  Castilla--La Mancha''.

%%%%%%%%%%%%%%%%%%%%%%%%%%%%%%%%%%%%%%%%%%%
\begin{figure}[ht]
\begin{center}
  \hspace{0mm}
  \psfig{file=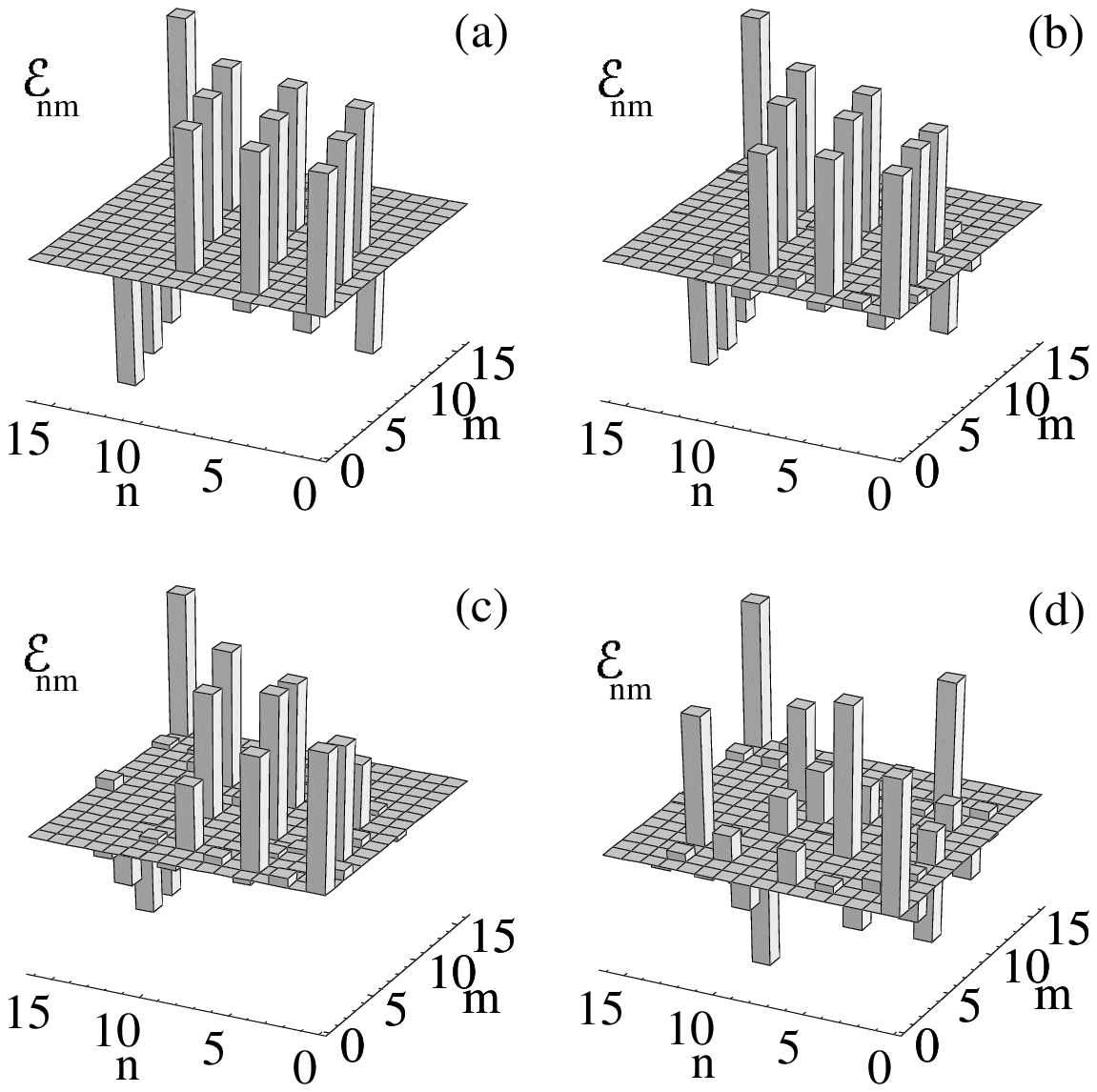,width=6.5cm}\\[0.3cm]
  \begin{caption}
{\sf ${\cal E}_{n,m.}$ in arbitrary units (see text for explanation) 
for a two--qubit gate in the ion trap quantum computer: (a) ideal
gate; (b,c,d) numerical simulation with the following parameters:
$\eta_{\rm cm}=\eta_{\rm r}=0.5$, $\Delta_{1}=\Delta_{2}=-\nu$, 
and $\Omega_1=\Omega_2=0.1,0.2,0.5\nu$ (b,c,d)}.
  \end{caption}
\end{center}
\end{figure}
%%%%%%%%%%%%%%%%%%%%%%%%%%%%%%%%%%%%%%%%%%%%%%%%%%%%%%%%%%%%%%%%%%%%%%
%%%%%%%%%%%%%%%%%%%%%%%%%%%%%%%%%%%%%%%%%%%
\begin{figure}[ht]
\begin{center}
  \hspace{0mm}
  \psfig{file=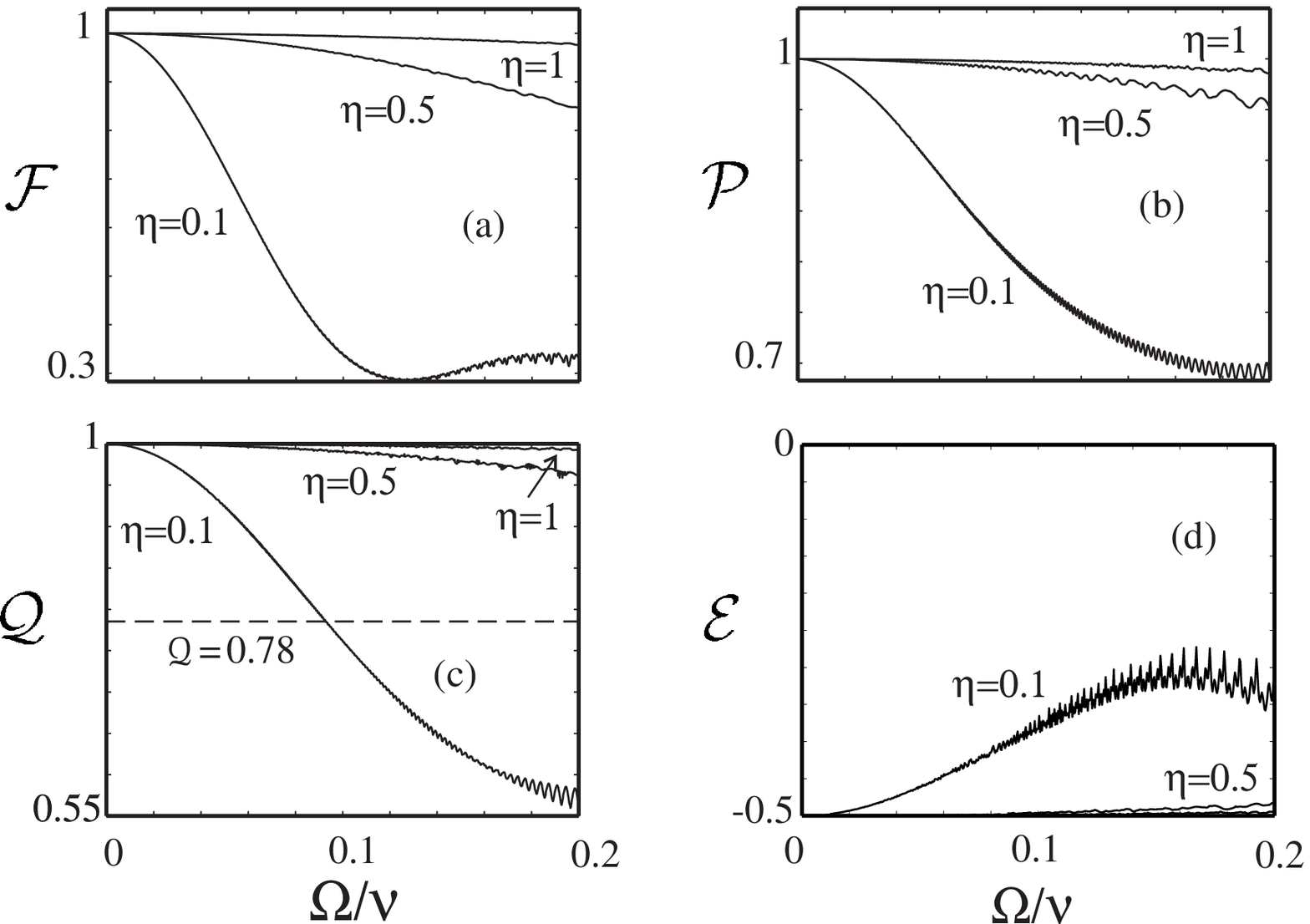,width=6.5cm}\\[0.3cm]
  \begin{caption}
{\sf Fidelity (a), Purity (b), Quantum Degree (c) and 
Entanglement Capability (d) for a two--qubit
gate in the ion trap quantum computer as a function of the Rabi
frequency. The values of the Lamb--Dicke parameter are indicated.
All the other parameters are as in Fig.~1}.
  \end{caption}
\end{center}
\end{figure}
%%%%%%%%%%%%%%%%%%%%%%%%%%%%%%%%%%%%%%%%%%%%%%%%%%%%%%%%%%%%%%%%%%%%%%


\begin{references}
\bibitem{note1}  In some gate models auxiliary
internal atomic levels are used, and the
output states can have amplitudes in those levels ($M>N=3$). 
In Eq.~(\ref{Woot}), $\hat \rho_{\rm out}$ denotes a projection of the 
density operator on ${\cal H}_2(1) \otimes {\cal H}_2(2)$, i.e. ${\rm Tr} 
\hat \rho _{\rm out} < 1$ which is the probability of ending up in the 
right Hilbert space.
\bibitem{note2}  In contrast to Ref.~\cite{Ci95} we have chosen 
lasers in a traveling wave configuration.
\bibitem{note3}  In Ref.~\cite{Ci95}, the Lamb--Dicke limit 
$\eta _{{\rm cm}}\ll 1$ was assumed. One can extend the
results beyond the Lamb--Dicke limit by correcting the time duration of the
laser pulses. Here we have used these corrected times.
\end{references}
\end{document}